\documentclass[journal]{IEEEtran}
\IEEEoverridecommandlockouts
\usepackage{cite}
\usepackage{amsmath,amssymb,amsfonts}
\usepackage[ruled]{algorithm2e}
\usepackage{graphicx}
\usepackage{textcomp}
\usepackage{xcolor}
\usepackage{color}
\usepackage{setspace}
\usepackage{subfigure}
\usepackage{caption}

\def\BibTeX{{\rm B\kern-.05em{\sc i\kern-.025em b}\kern-.08em
    T\kern-.1667em\lower.7ex\hbox{E}\kern-.125emX}}
\begin{document}
 
\title{Energy Allocation for Multi-User Cooperative Molecular Communication Systems in the Internet of Bio-Nano Things%
\thanks{This work was supported by China Postdoctoral Science Foundation under Grant 2023M732877.}%
\thanks{Dongliang Jing is with the College of Mechanical and Electronic Engineering, Northwest A\&F University, Shaanxi, China (dljing@nwafu.edu.cn).}%
\thanks{Lin Lin is with the College of Electronics and Information Engineering, Tongji University, Shanghai, China, (fxlinlin@tongji.edu.cn).}%
\thanks{Andrew W. Eckford is with the Department of Electrical Engineering and Computer Science, York University, Toronto, Ontario, Canada (aeckford@yorku.ca).}%
}
\author{
\IEEEauthorblockN{Dongliang Jing, Lin Lin, \IEEEmembership{Senior Member,~IEEE}, Andrew W. Eckford, \IEEEmembership{Senior Member,~IEEE}}
%
}


\markboth{To appear in IEEE Internet of Things Journal}%
{Shell \MakeLowercase{\textit{et al.}}: A Sample Article Using IEEEtran.cls for IEEE Journals}


\maketitle
\begin{abstract}
Cooperative molecular communication (MC) is a promising technology for facilitating communication between nanomachines in the Internet of Bio-Nano Things (IoBNT) field. However, the performance of IoBNT is limited by the availability of energy for cooperative MC. This paper presents a novel transmitter design scheme that utilizes molecule movement between reservoirs, creating concentration differences through the consumption of free energy, and encoding information on molecule types. The performance of the transmitter is primarily influenced by energy costs, which directly impact the overall IoBNT system performance. To address this, the paper focuses on optimizing energy allocation in cooperative MC for enhanced transmitter performance. Theoretical analysis is conducted for two transmitters. For scenarios with more than two users, a genetic algorithm is employed in the energy allocation to minimize the total bit error rate (BER). 
 Finally, numerical results show the effectiveness of the proposed energy allocation strategies in the considered cooperative MC system. 
\end{abstract}


\section{Introduction}
\IEEEPARstart{I}{nternet} of Bio-Nano Things (IoBNT) is a revolutionary paradigm that merges nanotechnology, biotechnology, and information technology to enable communication and coordination among bio-nanomachines (bio-NMs) \cite{akyildiz2015internet,kuscu2015internet,aghababaiyan2022enhanced,10065592}. These bio-NMs, also known as nanorobots or bio-nano things, are small devices made of biological or biocompatible materials. They have the potential to perform tasks such as computation, actuation, and sensing, opening up new avenues for applications like targeted drug delivery and healthcare monitoring \cite{ali2015internet,9793850, 9893201,chouhan2023interfacing}.

Communication among bio-NMs is a crucial aspect of IoBNT. Traditional electromagnetic communication methods are not suitable for the nanoscale due to the limited size of the bio-NMs and the unique environment they operate in. This has led to the exploration of alternative communication mechanisms, with molecular communication (MC) emerging as a promising solution. MC utilizes biochemical molecules as carriers of information, allowing bio-NMs to exchange messages through chemical signaling \cite{tang2021molecular,yaylali2023channel}. This communication approach takes advantage of the inherent properties of biochemical molecules, such as diffusion and reaction kinetics, to achieve communication at the nanoscale. By encoding information on the concentration of molecules \cite{kuran2011modulation}, types of molecules \cite{kuran2012interference}, or release timing \cite{eckford2007nanoscale}, MC enables bio-NMs to communicate and coordinate their actions, which is called cooperative MC. Cooperative MC enlarges the communication range efficiently and can perform more complex and challenging activities.

In cooperative MC, the information molecules are released by the transmitter and propagated to the receiver by diffusion or active transport methods. In particular, in diffusion-based cooperative MC, the released molecules undergo free diffusion in the fluid environment, following Brownian motion, without the need for any infrastructure or additional energy consumption \cite{fang2016distributed, tavakkoli2017optimal}. Research in cooperative MC has explored various aspects to enhance its performance and applications. In \cite{mosayebi2017cooperative}, collaborative abnormality detection using cooperative MC was studied, considering multi-sensors and a fusion center. The performance of a cooperative MC system within a cylindrical-shaped channel was analyzed in \cite{dhok2021cooperative}. In \cite{fang2016distributed}, global decision-making process in a cooperative MC system was studied, in which distributed receivers collaboratively determine a transmitter's signal, leading to enhanced reliability. On the other hand, in \cite{fang2019symbol}, a symbol-by-symbol maximum likelihood detection scheme was proposed for cooperative MC, involving multiple receivers, a fusion center, and decision-making based on observations from all receivers. Channel characterization with multiple fully absorbing receivers was analyzed considering the neighboring interference \cite{ferrari2022channel}. Considering multiple fully absorbing
receivers, in \cite{sabu2022channel}, the authors analyzed the mathematical expression for the hitting probability of a molecule emitted from a point source on each fully absorbing receiver. 

Of particular relevance to this paper are studies on constrained resources and intersymbol interference. 
In terms of constrained molecule resources in cooperative MC, the optimization of molecular resource allocation, including the assignment of the types of molecules and the number of molecules of a type, was studied in \cite{chen2021resource}. To further improve cooperative MC performance, the unavoidable multi-user interference arising from a large number of bio-NMs in the medium was addressed by considering the direction of releasing molecules as a new property \cite{aghababaiyan2022enhanced}. Further research in this direction includes \cite{vakilipoor2023localizing}, in which localization of the unknown receiver in the cooperative MC was studied by mapping the likelihood of the position of the unknown receiver from the known receivers’ perspective. In \cite{miao2022levenberg}, a cooperative source
localization method employing multiple spherical absorption receivers is proposed for MC.

In MC, a reservoir is essential to store the information molecules \cite{farsad2016comprehensive,kuscu2019transmitter}. The information molecules may be either synthesized by the transmitter or harvested from the environment \cite{demiray2013direct,nakano2013molecular}. The paper \cite{ahmadzadeh2022molecule} investigates mathematical models for molecule-harvesting transmitters in which molecules are recaptured upon contact with the transmitter's harvesting units. The method of genetic engineering by modifying the genes of cells to generate desired molecules was studied in \cite{unluturk2015genetically}, however, usually only one kind of molecule is generated, ruling out schemes such as molecule shift keying.
Harvesting molecules from the environment generates multiple types of molecules, however, the molecules are mixed and require separation into purified reservoirs.
In \cite{eckford2018thermodynamic}, the purification process of moving molecules from one reservoir to the other was studied, and the creation of the reservoirs was found to consume free energy. However, for reasonable energy costs, different types of molecules can not be totally separated, resulting in partially impure reservoirs, leading to severe interference at the receiver.  In \cite{chude2015diffusion}, the transmitter was modeled as a spherical structure, and molecules were produced at its center. In \cite{huang2021membrane}, the errors that arise from the molecules being probabilistically released from the transmitter membrane was studied. In \cite{9785829}, limited storage capacity at the transmitter was considered, where the transmitter released a part of its stored molecules to encode bits, and filled the storage up to the maximum storage capacity in the remaining time. 

In this paper, we explore a multi-user cooperative molecular communication system where the total amount of available free energy is constrained for a given size \cite{alarcon2016nanomotors, novotny2020nanorobots}. Information molecules are collected from the environment, comprising a mixture of different types of molecules. Building on the work presented in \cite{eckford2018thermodynamic}, we focus on two species of molecules within two reservoirs. The concentration difference between these reservoirs is established by actively moving molecules between them, a process that consumes energy. Consequently, the allocation of energy among the users (transmitters) becomes crucial, as it directly impacts the overall system performance. In light of the limited energy available within the specified area, our study addresses the proper distribution of energy among the users based on their respective sizes. The objective is to achieve a lower bit error rate (BER) at the transmitters while considering the constraints posed by the available energy resources. By optimizing the energy allocation strategy, we aim to enhance the efficiency and reliability of the communication system. Unlike \cite{eckford2018thermodynamic}, which primarily addresses energy consumption during molecule movement between reservoirs, this paper focuses on optimizing energy allocation in cooperative MC to enhance transmitter performance, considering the imperfect transmitter created through molecule transfer between reservoirs.
The potential application of energy allocation in MC is critical for addressing energy constraints, especially in imperfect transmitter MC, where the creation of transmitters consumes energy. This is particularly relevant in health monitoring applications. Due to the small size of multiple bio-NMs, the available resources are limited. To enable the transmitters to operate more effectively, bio-NMs must collect a mixture of molecules from the environment, which consumes free energy. However, the total amount of available free energy is constrained, considering a specific size. The performance of the transmitters primarily relies on the allocated energy. Therefore, it is essential to allocate energy properly among the bio-NMs to enhance the performance of the transmitters.

The main contributions of this paper are summarized as follows.
\begin{itemize}
\item We present a novel cooperative MC system that incorporates imperfect transmitters. In this system, a specific type of molecule is actively moved from one reservoir to another to create the transmitter, and we analyze both the energy consumption involved and the performance of the resulting transmitter.
\item Due to the fact that in the considered cooperative MC system, the total energy is constrained. To achieve the best performance of cooperative MC, energy allocation strategies are studied.
\item Monte Carlo simulations are conducted to evaluate the effectiveness of the proposed energy allocation strategies. 
\end{itemize}

The rest of this paper is organized as follows. In section II, we introduce the problem statement in the considered cooperative MC system. In section III, we perform optimization and analyze the resulting optimization method. In section IV, numerical and simulation results are presented. Finally, in Section V, we conclude this paper.

\section{Problem Statement}
\subsection{System Model}
\begin{figure}[!h]
  \centering
  \includegraphics[width=0.5\textwidth]{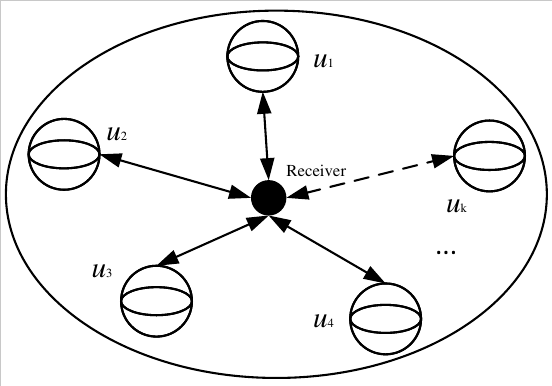}
  \caption{A depiction of the cooperative communication system with the multi-user scenario. The users communicate with the receiver by releasing different types of molecules.} \label{system_model}
\end{figure}

In this paper, we consider a cooperative MC system, similar to that presented in \cite{chen2021resource,9828497}, which consists of a spherical passive receiver and $K$ point users. In this paper, we focus on the uplink communication process, therefore, there is one receiver and $K$ transmitters. As depicted in Fig. \ref{system_model}, $K$ users communicate with a central receiver nanomachine by MC. In this cooperative MC system, 
we assume that the users are able to collect information molecules from the environment to its reservoirs or store information molecules in its reservoirs. In this paper, we explore the use of MoSK, where two distinct types of molecules are considered, and these molecules are collected and stored in their respective reservoirs. The molecular types between different users can be the same or different. For the same types of molecules among the users, a time of release modulation scheme can be employed, allowing information to be encoded in the timing of the release of information molecules, then, the receiver can distinguish the originating transmitter responsible for the molecule release based on the pre-set transmission bit interval. For the different types of molecules among the users, the information can be encoded in the types of information molecules, then, at the receiver, receptors are capable of selectively binding to specific ligand structures, enabling the distinction of the originating transmitter responsible for the molecule release. However, it is important to note that this paper does not delve into the specifics of molecule types among different users. 

\begin{figure}[!h]
  \centering
  \includegraphics[width=0.5\textwidth]{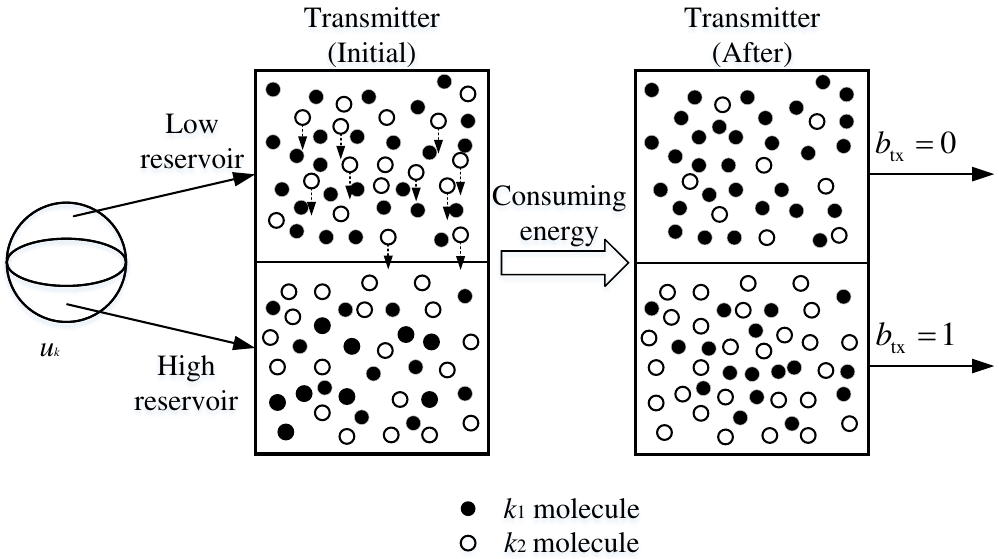}\\
  \caption{A depiction of the imperfect transmitter. $k_1$ molecules are filled circles, while $k_2$ molecules are unfilled circles. Both the low reservoir and high reservoir are filled with $k_1$ and $k_2$ molecules. The model is inspired by the Maxwell's Demon thought experiment \cite{leff2002maxwell}.}\label{imperfect_transmitter}
\end{figure}

Unlike the ideal transmitter, in the considered cooperative MC system, we focus on an imperfect transmitter, which means the collected or stored molecules are mixed. As shown in Fig. \ref{imperfect_transmitter}, the imperfect transmitter is composed of two reservoirs, namely low reservoir and high reservoir, in each reservoir, there are different two types of molecules, for example, in the $U_k$ transmitter, there are fulfilled with $k_1$ and $k_2$ molecules. And the initial state of molecules in the low and high reservoirs are the same, which means the initial concentration of $k_1$ molecules are the same in the low and high reservoir, equally applicable to $k_2$ molecules. Then, to create the transmitter, $k_2$ molecules are moved from the low reservoir to the high reservoir by consuming free energy. 
As the total energy in the cooperative MC is limited, therefore, it should be properly allocated to achieve the best performance.

Creating the reservoir requires an input of free energy because of the changing chemical potential. From \cite{eckford2018thermodynamic}, for the $U_k$ transmitter, if $m$ molecules of type $k_2$ are moved from the low reservoir to the high reservoir (where $m$ is much smaller compared to the total number of molecules in the reservoir), then the energy cost can be expressed as
\cite[Eqn. (25)]{eckford2018thermodynamic} 
\begin{align}
\begin{split}
\label{equ3}
E_{u_k} =& {n_{k,H}}KT\left[  - \left( {1 + \frac{n_{k,L}}{n_{k,H}}} \right)c_k\log c_k + {c_{k,H}}\log {c_{k,H}} \right.\\
&\left.+ \frac{n_{k,L}}{n_{k,H}}{c_{k,L}}\log {c_{k,L}} \right],   
\end{split}
\end{align}
where $n_{k,L}$ and $n_{k,H}$ are the total number of molecules in the low and high reservoirs in the $U_k$ transmitter, respectively,  $K$ is Boltzmann's constant, $T$ is the absolute temperature, $c_k$ is the initial concentration of $k_2$ molecules in the reservoirs, $c_{k,L}$ and $c_{k,H}$ are the fraction of $k_2$ molecules in the low and high reservoirs in the $U_k$ transmitter, respectively. Then, after $m$ $k_2$ molecules are moved from the low reservoir to the high reservoir, the mole fractions of type $k_2$ molecules in the low and high reservoirs can be expressed as
\begin{align}
\label{equ2cl}
\begin{split}
{c_{k,L}} = c_k - \frac{m}{{{n_{k,L}}}},
\end{split}
\end{align}
and 
\begin{align}
\label{equ2ch}
\begin{split}
{c_{k,H}} = c_k + \frac{m}{{{n_{k,H}}}},
\end{split}
\end{align}
{\color{black}respectively.}

To analyze the performance of the transmission scheme, from (\ref{equ3}), 
we focus on the number of moved molecules $m$ varies with the energy cost $E_k$.
Assume that $n_{k,L}=n_{k,H}$, and let $n_k = n_{k,L}+n_{k,H}$ represent the total number of molecules in the $k$ user. Then (\ref{equ3}) can be expressed as
\begin{align}
\label{equ4}
\begin{split}
E_{u_k}  &= {n_{k,H}}kT\left[ \left( {c_k + \frac{{2m}}{n_k}} \right)\log \left( {1 + \frac{{2m}}{{c_kn_k}}} \right) \right. \\
&+\left. \left( {c_k - \frac{{2m}}{n_k}} \right)\log \left( {1 - \frac{{2m}}{{c_kn_k}}} \right) \right].
\end{split}  
\end{align}

To simplify the notation, define $\alpha = \frac{{2m}}{n_k}$, the number of moved molecules as a fraction of the reservoir size, and define
$\beta  = \frac{\alpha }{c_k} = \frac{{2m}}{{c_kn_k}}$. Then
(\ref{equ4}) can be expressed as
\begin{align}
\label{equ5}
E_{u_k} = {n_{k,H}}KT\left[ {\left( {c_k + \alpha } \right)\log \left( {1 + \beta } \right) + \left( {c_k - \alpha } \right)\log \left( {1 - \beta } \right)} \right].    
\end{align}
By assumption, the number of moved molecules $m$ is much smaller than the number of molecules in the low or high reservoir. Thus, $0 < \beta \ll 1$. By employing a Taylor series \cite{5273711,5264196}, we can write
\begin{align}
\label{equ6}
\begin{split}
\log \left( {1 + \beta } \right) = \beta  - \frac{1}{2}{\beta ^2} + o\left( {{\beta ^3}} \right), \\
\log \left( {1 - \beta } \right) =  - \beta  - \frac{1}{2}{\beta ^2} + o\left( {{\beta ^3}} \right).
\end{split}
\end{align}
Therefore, (\ref{equ5}) can be expressed as
\begin{align}
\label{equ7}
\begin{split}
E_{u_k} \!\!&=\! {n_{k,H}}KT\left[ {\left( {c_k \!+\! \alpha } \right)\left( {\beta  - \frac{1}{2}{\beta ^2}} \right) \!+\! \left( {c_k \!-\! \alpha } \right)\left( { - \beta  - \frac{1}{2}{\beta ^2}} \right)} \right] \\ 
&+ o\left( {{\beta ^3}} \right)\\ 
 &= {n_{k,H}}KT\left[ { - c_k{\beta ^2} + 2\alpha \beta } \right] + o\left( {{\beta ^3}} \right)\\ 
 &= {n_{k,H}}KT\left[ { - c{{\left( {\frac{{2m}}{{c_kn_k}}} \right)}^2} + 2\frac{{2m}}{n_k}\left( {\frac{{2m}}{{c_kn_k}}} \right)} \right] + o\left( {{\beta ^3}} \right)\\
 &= 2KT\frac{{{m^2}}}{{c_k{n_k}}} + o\left( {{\beta ^3}} \right).
\end{split}
\end{align}
Neglecting the higher-order infinitesimal term $o\left( {{\beta ^3}} \right)$, for the given energy cost, the number of moved molecules $m$ can be expressed as
\begin{align}
\label{equ8}
m = \sqrt {\frac{{c_kn_k}}{{2KT}}E_{u_k}}.
\end{align}
Then, based on (\ref{equ2cl}) and (\ref{equ2ch}), $c_{k,L}$ and $c_{k,H}$ can be expressed as
\begin{align}
\begin{split}
\label{moved_molecules_energy_ch51}
{c_{k,L}} = c_k - \sqrt {\frac{{c_k}}{{KT{n_{k,L}}}}E_{u_k}},   
\end{split}    
\end{align}
and 
\begin{align}
\begin{split}
\label{moved_molecules_energy_ch52}
{c_{k,H}} = c_k + \sqrt {\frac{{c_k}}{{KT{n_{k,H}}}}E_{u_k}},    
\end{split}    
\end{align}
{\color{black}respectively.}

{\color{black}It is illustrated by (\ref{moved_molecules_energy_ch51}) and (\ref{moved_molecules_energy_ch52}) that}, within the considered imperfect transmitter model, the number of moved molecules, $m$, is influenced not only by the consumed energy $E_{u_k}$, but also by the number of molecules in the reservoirs, and the initial concentration of $k_2$ molecules $c_k$.

In the communication process, as shown in Fig. \ref{imperfect_transmitter}, molecular shift keying (MoSK) is employed, which means, during a bit interval, to transmit bit 0, $N_{tx}$ molecules are released from the low reservoir, and transmit bit 1, $N_{tx}$ molecules are released from the high reservoir. It is important to note that the molecules released from each reservoir are selected at random to ensure an unbiased and representative sample. Compared to the distance between the transmitter and the receiver, the size of the information molecules and reservoirs are assumed to be very small. Thus, the transmitter can be modeled by a point transmitter.

As noted above, in the communication process, to transmit bit 0, a large number of molecules $N_m$ are selected from the low reservoir, and this selection is taken uniformly at random from the reservoir. In this selection, the average number of $k_2$ molecules is a Bernoulli random variable with $N_m$ trials and success probability $c_{k,L}$.
Similarly, to transmit bit 1, a larger number of molecules $N_m$ are selected from the high reservoir, the number of $k_2$ molecules in this selection is a Bernoulli random variable with $N_m$ trials and success probability $c_{k,H}$.

As in the reservoirs, the initial concentration of $k_2$ molecules is $c_k$ and $k_1$ molecules is $1-c_k$. Therefore, the initial concentration ratio of $k_1$ and $k_2$ molecules is ${c_{k_1,L}}/{c_{k_2,L}} = \frac{1}{c_k}-1$. After moving $m$ $k_2$ molecules from the low reservoir to the high reservoir, the concentration ratio of $k_1$ and $k_2$ molecules in the low reservoir should be ${c_{k_1,L}}/{c_{k_2,L}} \geq \frac{1}{c_k}-1$. And the concentration ratio of $k_1$ and $k_2$ molecules in the high reservoir should be ${c_{k_1,H}}/{c_{k_2,H}} < \frac{1}{c_k}-1$. Therefore, in a time slot, the bit information represented by the information molecules released by the transmitter is determined by
\begin{align}
\begin{split}
\label{b_tx_k}
{b_{tx,u_k}} = \left\{ {\begin{array}{*{20}{c}}
{0,}&{{{{c_{k_1,L}}} \mathord{\left/
 {\vphantom {{{c_{k_1,L}}} {{c_{k_2,L}}}}} \right.
 \kern-\nulldelimiterspace} {{c_{k_2,L}}}} \geq \frac{1}{c_k} - 1,} \vspace{1.5ex}\\ 
{1,}&{{{{c_{k_1,H}}} \mathord{\left/
 {\vphantom {{{c_{k_2,H}}} {{c_{k_1,H}}}}} \right.
 \kern-\nulldelimiterspace} {{c_{k_2,H}}}} < \frac{1}{c_k} - 1,}
\end{array}} \right.
\end{split}
\end{align}
where $b_{tx,u_k}$ is the transmitted bit for the $u_k$ transmitter.

To transmit bit $0$, $N_m$ molecules are selected from the low reservoir and released, then, according to (\ref{b_tx_k}), the probability that the concentration of released molecules satisfied ${c_{k_1, L}}/{c_{k_2, L}}\geq\frac{1}{c_k}-1$ can be expressed as 
\begin{equation}
\label{equ10}
{P_k}\left( {Y = 0|X = 0} \right) = \frac{{\sum\limits_{i = \left\lfloor {N_m(1-c_k)} \right\rfloor  + 1}^{{N_m}} {C_{{n_{k,L}}\left( {1 - {c_{{k_2},L}}} \right)}^iC_{{n_{k,L}}{c_{k_2,L}}}^{{N_m} - i}} }}{{C_{{n_{k,L}}}^{{N_m}}}}.
\end{equation}
In order to compare the concentration of $k_1$ molecules and $k_2$ molecules, $N_m$ takes an odd number. Of the $N_m$ molecules, the selected number of molecules of type $k_1$, i.e. $i$, can be expressed as  $ {{\left( {C_{{n_{k,L}}\left( {1 - {c_{k_2,L}}} \right)}^iC_{{n_{k,L}}{c_{k_2,L}}}^{{N_m} - i}} \right)} \mathord{\left/ {\vphantom {{\left( {C_{{k,n_L}\left( {1 - {c_{k_2,L}}} \right)}^iC_{{n_{k,L}}{c_{k_2,L}}}^{{N_m} - i}} \right)} {C_{{n_{k,L}}}^{{N_m}}}}} \right.} {C_{{n_{k,L}}}^{{N_m}}}}$  and follows the Hypergeometric distribution. As $n_{k, L}$ is large, the Hypergeometric distribution can be approximated by the Binomial distribution, ${X_0}\left( i \right) \sim B\left( {i,1 - {c_{k_2,L}}} \right)$ \cite{chattamvelli2020discrete}.
Also considering $N_m$ is large, the Binomial distribution can be approximated by Normal distribution ${X_0} \sim N\left( {{\mu _0},\sigma _0^2} \right)$, where
\begin{align}
\label{equ11}
\begin{gathered}
  {\mu _0} = {N_m}(1-{c_{k_2,L}}), \hfill \\
  \sigma _0^2 = {N_m}{c_{k_2,L}}\left( {1 - {c_{k_2,L}}} \right). \hfill \\ 
\end{gathered}    
\end{align}
Therefore, for the $k$th transmitter, the probability of transmit bit 0 ($b_{tx,u_k}=0$) and the transmitted molecules satisfied ${c_{k_1,L}}/{c_{k_2,L}}\geq\frac{1}{c_k}-1$ can be expressed as
\begin{align}
\begin{split}    
\label{X0Y0}
{P_{k}}\left( {Y = 0|X = 0} \right) &= \Phi \left( {\frac{{{N_m} - \mu_0 }}{{{\sigma _0}}}} \right)\\ 
&- \Phi \left( {\frac{{\left\lfloor N_m(1-c_k) \right\rfloor  + 1 - \mu_0 }}{{{\sigma _0}}}} \right) , 
\end{split}
\end{align}
where 
$\Phi \left( x \right) = \frac{1}{{\sqrt {2\pi } }}\int_{ - \infty }^x {{e^{ - \frac{{{t^2}}}{2}}}} dt$ is the cumulative distribution function of the standard normal distribution. In the considered MC, it is reasonable to assume that the quantity $N_m(1-c_k)$ is sufficiently large, then it is possible to approximate ${\left\lfloor N_m(1-c_k) \right\rfloor + 1 - \mu_0 }$ as $N_m(1-c_k)- \mu_0$. Then, (\ref{X0Y0}) can be expressed as
\begin{align}
\begin{split}    
\label{X0Y0v2}
{P_{k}}\left( {Y = 0|X = 0} \right) &= \Phi \left( {\frac{{{N_m} - \mu_0 }}{{{\sigma _0}}}} \right)\\ 
&- \Phi \left( {\frac{{ N_m(1-c_k)- \mu_0 }}{{{\sigma _0}}}} \right). 
\end{split}
\end{align}

%

Similarly, the probability of transmitting bit 1, $N_m$ molecules are selected from the high reservoir and released, then, according to (\ref{b_tx_k}), the probability that the concentration of released molecules satisfied ${c_{k_1,H}}/{c_{k_2,H}}<\frac{1}{c_k}-1$ can be expressed as 
\begin{equation}
\label{equ13}
{P_k}\left( {Y = 1|X = 1} \right) = \frac{{\sum\limits_{i = \left\lfloor N_mc_k \right\rfloor  + 1}^{{N_m}} {C_{{n_{k,H}}{c_{k_2,H}}}^iC_{{n_{k,H}}\left( {1 - {c_{k_2,H}}} \right)}^{{N_m} - i}} }}{{C_{{n_{k,H}}}^{{N_m}}}}.
\end{equation}

Similarly, for selecting $N_m$ molecules from the high reservoir, $n_L$ is large and $N_m$ is large, then the probability of $i$ $k_2$ molecules follows Normal distribution ${X_1} \sim N\left( {{\mu _1},\sigma _1^2} \right)$, where
\begin{align}
\label{equ14}
\begin{gathered}
  {\mu _1} = {N_m}{c_{k_2,H}}, \hfill \\
  \sigma _1^2 = {N_m}{c_{k_2,H}}\left( {1 - {c_{k_2,H}}} \right). \hfill \\ 
\end{gathered}     
\end{align}
Therefore, for the $k$th transmitter, the probability of transmitting bit 1 and the probability that the released molecules satisfied ${c_{k_1,H}}/{c_{k_2,H}}<\frac{1}{c_k}-1$ can be expressed as 
\begin{align}
\begin{split}
\label{X1Y1}
{P_{k}}\left( {Y = 1|X = 1} \right) =& \Phi \left( {\frac{{{N_m} - \mu_1 }}{{{\sigma _1}}}} \right) \\
&- \Phi \left( {\frac{{\left\lfloor N_mc_k \right\rfloor  + 1 - \mu_1 }}{{{\sigma _1}}}} \right) .  
\end{split}
\end{align}
In the considered MC, assuming that $N_mc_k$ is sufficiently large, it becomes feasible to approximate ${\left\lfloor N_m-c_k\right\rfloor + 1 - \mu_1 }$ as $N_mc_k- \mu_1$. Then, (\ref{X1Y1}) can be expressed as
\begin{align}
\begin{split}
\label{X1Y1v2}
{P_{k}}\left( {Y = 1|X = 1} \right) =& \Phi \left( {\frac{{{N_m} - \mu_1 }}{{{\sigma _1}}}} \right) \\
&- \Phi \left( {\frac{{N_mc_k -\mu_1 }}{{{\sigma _1}}}} \right) .  
\end{split}
\end{align}
%
%

Assuming that each bit $\{0,1\}$ is equiprobable at the transmitter, the BER raised by the $k$th transmitter during a time slot can be expressed as
\begin{align}
\begin{split}
\label{ber_any}
{P_{e,u_k}} =& \frac{1}{2}\left[ \left( {1 - {P_{k}}\left( {Y = 0|X = 0} \right)} \right) \right.\\
&+ \left. \left( {1 - {P_{k}}\left( {Y = 1|X = 1} \right)} \right) \right].
\end{split}
\end{align}
where ${P_{k}}\left( {Y = 0|X = 0} \right)$ and ${P_{k}}\left( {Y = 1|X = 1} \right)$ {\color{black}are given by (\ref{X0Y0v2}) and (\ref{X1Y1v2}), respectively.} 

\subsection{Problem Formulation}
For a multi-user cooperative MC system with predetermined transmitter parameters, the efficiency of the transmitters hinges on energy allocation. In this study, we delve into the optimization of energy allocation with the aim of maximizing transmitter performance. The central optimization objective involves minimizing the cumulative BER across all users. 
In this paper, we address the problem of minimizing the total BER in a multi-user cooperative MC system. The primary goal is to determine an energy allocation configuration ${E_{u_1}}, {E_{u_2}}, ..., {E_{u_k}}$ that minimizes the aggregate total BER while adhering to specified constraints. The problem can be formulated as follows:

\begin{align}
\begin{split}
\label{optimazation_problem}
&\min_{{E_{u_1}}, {E_{u_2}}, ..., {E_{u_k}}} \sum_{k = 1}^K {P_{e,u_k}}, \\
\text{subject to} \quad &\sum_{k=1}^K E_{u_k} \leq E_{\text{total}}, \\
&P_{e,k} \leq \text{BER}_{\text{threshold}}, \quad \forall k = 1, 2, \ldots, K, \\
&E_{u_k} \geq 0, \quad \forall k = 1, 2, \ldots, K.
\end{split}
\end{align}
Here, $\sum_{k=1}^K E_{u_k} \leq E_{\text{total}}$ represents the constraint that the total energy allocated is less than the total energy $E_{\text{total}}$. By imposing the constraint $P_{e,k} \leq \text{BER}_{\text{threshold}}$, we ensure that the BER of each link, denoted as $P_{e,k}$, does not exceed the predetermined upper limit, defined as $\text{BER}_{\text{threshold}}$. This constraint serves as a safeguard against the BER of a link increasing to unacceptably high levels, which could potentially result in a communication breakdown on that specific link. The constraint $E_{u_k} \geq 0, \quad \forall k = 1, 2, \ldots, K$ guarantees that all transmitters are allocated a non-negative amount of energy and prevents the energy from being allocated to only a fraction of transmitters.

\section{Optimization Analysis}
In this section, we study the optimization for allocating energy among the transmitters to minimize the total BER in a multi-user cooperative MC system. First, we take $K=2$ as an example to illustrate the principles of energy allocation. Meanwhile, when $K>2$, we employ the genetic algorithm (GA) to analyze and solve the optimization problem of energy allocation with the objective of minimizing the total BER. 
\subsection{Optimization scheme for $K=2$}
In this subsection, we study the energy allocation strategy to achieve the minimization of the total BER in a multi-user communication system. According to the aforementioned, we can rewrite (\ref{optimazation_problem}) for $K=2$ as
\begin{align}
\begin{split}
\label{optimazation_problem_v2}
&\min_{\rho}\{{P_{e,u_1}+{P_{e,u_2}}}\}, \\
\text{subject to} \quad &E_{u_1} + E_{u_2} \leq \rho E + (1-\rho) E = E,\\
& 0 <\rho < 1,
\end{split}
\end{align}
where $\rho E$ and $(1-\rho) E$ are the energy allocated to users $u_1$ and $u_2$, respectively, and $\rho$ is the allocation coefficient. 

We denote
\begin{align}
\begin{split}
\label{f_rho}
f(\rho) = {P_{e,u_1}+{P_{e,u_2}}},
\end{split}
\end{align}
where $P_{e,u_1}$ and $P_{e,u_2}$ are derived in the Appendix. 

Assuming the same number of molecules in the low and high reservoirs, namely $n_{k, L} = n_{k, H} = \frac{1}{2}n_k$, then, for simplifying the donation, we denote
\begin{align}
\begin{split}
\psi  = \frac{{{c_k}}}{{KT{n_{k,L}}}} = \frac{{{c_k}}}{{KT{n_{k,H}}}}. \\
\end{split}
\end{align}

For simplification, we consider the initial concentration $c_k=1/2$, signifying an equal number of molecules of types $k_1$ and $k_2$ within the reservoirs. Then, $P_{e,u_1}$ and $P_{e,u_2}$ can be expressed as
\begin{align}
\begin{split}
{P_{e,{u_1}}} &= 1 - \Phi \left( {\sqrt {\frac{{{N_m}\left( {\frac{1}{2} - \sqrt {\psi \rho E} } \right)}}{{\frac{1}{2} + \sqrt {\psi \rho E} }}} } \right) \\
&+ \Phi \left( {\frac{{ - \sqrt {{N_m}\psi \rho E} }}{{\sqrt {\left( {\frac{1}{2} - \sqrt {\psi \rho E} } \right)\left( {\frac{1}{2} + \sqrt {\psi \rho E} } \right)} }}} \right) \\
&=1 - \Phi \left( {\sqrt {\frac{{{N_m}\left( {\frac{1}{2} - \sqrt {\psi \rho E} } \right)}}{{\frac{1}{2} + \sqrt {\psi \rho E} }}} } \right) \\
&+ \Phi \left( { - \sqrt {\frac{{{N_m}\psi \rho E}}{{\frac{1}{4} - \psi \rho E}}} } \right),
\end{split}
\end{align}
and
\begin{align}
\begin{split}
{P_{e,{u_2}}} &= 
 1 - \Phi \left( {\sqrt {\frac{{{N_m}\left( {\frac{1}{2} - \sqrt {\psi \left( {1 - \rho } \right)E} } \right)}}{{\frac{1}{2} + \sqrt {\psi \left( {1 - \rho } \right)E} }}} } \right) \\
&+ \Phi \left( -\sqrt{\frac{  {N_m}\psi \left( {1 - \rho } \right)E }{\frac{1}{4} - \psi \left( {1 - \rho } \right)E}} \right),   
\end{split}
\end{align}
respectively.







We denote $\Phi_1 = \Phi \left( {\sqrt {\frac{{{N_m}\left( {\frac{1}{2} - \sqrt {\psi \rho E} } \right)}}{{\frac{1}{2} + \sqrt {\psi \rho E} }}} } \right)$ and $\Phi_2 = \Phi \left( { - \sqrt {\frac{{{N_m}\psi \rho E}}{{\frac{1}{4} - \psi \rho E}}} } \right)$, then,

\begin{align}
\begin{split}
\frac{{\partial {P_{e,{u_1}}}}}{{\partial \rho }} =  - \frac{{\partial {\Phi _1}}}{{\partial \rho }} + \frac{{\partial {\Phi _2}}}{{\partial \rho }}.    
\end{split}    
\end{align}

Assuming $b = {\sqrt {\frac{{{N_m}\left( {\frac{1}{2} - \sqrt {\psi \rho E} } \right)}}{{\frac{1}{2} + \sqrt {\psi \rho E} }}} }$, we can obtain $\Phi_1 = \Phi \left( b \right)$. As a result, the partial derivatives,  $\frac{{\partial {\Phi _1}}}{{\partial \rho }}$ and $\frac{{\partial {\Phi _2}}}{{\partial \rho }}$ can be expressed as follows

\begin{align}
\begin{split}
\frac{{\partial {\Phi _1}}}{{\partial \rho }} &= \frac{{\partial \Phi_1 }}{{\partial b}} \times \frac{{\partial b}}{{\partial \rho }}\\
 &= -\phi \left( b \right) \times \frac{1}{{2b}} \times \frac{{{N_m}\psi E}}{{2\sqrt {\psi \rho E} {{\left( {\frac{1}{2} + \sqrt {\psi \rho E} } \right)}^2}}}\\
 & = - \frac{1}{4}\phi \left(\sqrt {\frac{{{N_m}(\frac{1}{2} - \sqrt {\psi \rho E} )}}{{\frac{1}{2} + \sqrt {\psi \rho E} }}} \right)\\
 &\times \frac{{\sqrt {{N_m}\psi E} }}{{\sqrt {\rho (\frac{1}{4} - \psi \rho E)}  \times (\frac{1}{2} + \sqrt {\psi \rho E} )}},
\end{split}
\end{align}
where $\phi\left( \right)$ the standard normal probability density function.

\begin{align}
\begin{split}
\frac{{\partial {\Phi _2}}}{{\partial \rho }} &=  - \phi \left( { - \sqrt {\frac{{{N_m}\psi \rho E}}{{\frac{1}{4} - \psi \rho E}}} } \right)\frac{1}{{2\sqrt {\frac{{{N_m}\psi \rho E}}{{\frac{1}{4} - \psi \rho E}}} }}\frac{{\frac{1}{4}{N_m}\psi E}}{{{{\left( {\frac{1}{4} - \psi \rho E} \right)}^2}}}  \\
&= - \frac{1}{8}\phi \left( - \sqrt {\frac{{{N_m}\psi \rho E}}{{\frac{1}{4} - \psi \rho E}}} \right) \times \frac{{\sqrt {{N_m}\psi E} }}{{\sqrt \rho  {{(\frac{1}{4} - \psi \rho E)}^{\frac{3}{2}}}}}.
\end{split}    
\end{align}

We denote $\Phi_3  = \Phi\left( {\sqrt {\frac{{{N_m}\left( {\frac{1}{2} - \sqrt {\psi \left( {1 - \rho } \right)E} } \right)}}{{\frac{1}{2} + \sqrt {\psi \left( {1 - \rho } \right)E} }}} } \right)$ and $\Phi_4 = \Phi \left( -\sqrt{\frac{  {N_m}\psi \left( {1 - \rho } \right)E }{\frac{1}{4} - \psi \left( {1 - \rho } \right)E}} \right)$, then,
\begin{align}
\begin{split}
\frac{{\partial {P_{e,{u_2}}}}}{{\partial \rho }} =  - \frac{{\partial {\Phi _3}}}{{\partial \rho }} + \frac{{\partial {\Phi _4}}}{{\partial \rho }}.    
\end{split}    
\end{align}
Assuming $e = {\sqrt {\frac{{{N_m}\left( {\frac{1}{2} - \sqrt {\psi \left( {1 - \rho } \right)E} } \right)}}{{\frac{1}{2} + \sqrt {\psi \left( {1 - \rho } \right)E} }}} }$, we can then obtain $\Phi_3 = \Phi \left( e \right)$. As a result, the partial derivatives $\frac{{\partial {\Phi _3}}}{{\partial \rho }}$ and $\frac{{\partial {\Phi _4}}}{{\partial \rho }}$
can be expressed as
\begin{align}
\begin{split}
\frac{{\partial {\Phi _3}}}{{\partial \rho }} &= \frac{{\partial {\Phi _3}}}{{\partial e}} \times \frac{{\partial e}}{{\partial \rho }}\\
 &= -\phi \left( e \right) \times \frac{1}{{2e}} \times \frac{{{N_m}\psi E}}{{2\sqrt {\psi \left( {1 - \rho } \right)E} {{\left( {\frac{1}{2} + \sqrt {\psi \left( {1 - \rho } \right)E} } \right)}^2}}}\\
 &=\frac{1}{4}\phi \left(\sqrt {\frac{{{N_m}(\frac{1}{2} - \sqrt {\psi (1 - \rho )E} )}}{{\frac{1}{2} + \sqrt {\psi (1 - \rho )E} }}} \right)\\
 &\times  \frac{{\sqrt {{N_m}\psi E} }}{{\sqrt {(1 - \rho )} \sqrt {\frac{1}{4} - \psi (1 - \rho )E} (\frac{1}{2} + \sqrt {\psi (1 - \rho )E} )}},
\end{split}
\end{align}
and
\begin{align}
\begin{split}
\frac{{\partial {\Phi _4}}}{{\partial \rho }} &=  \phi \left( { - \sqrt {\frac{{{N_m}\psi \left( {1 - \rho } \right) E}}{{\frac{1}{4} - \psi \left( {1 - \rho } \right) E}}} } \right)\frac{1}{{2\sqrt {\frac{{{N_m}\psi \left( {1 - \rho } \right) E}}{{\frac{1}{4} - \psi \left( {1 - \rho } \right) E}}} }} \\
&\times \frac{{\frac{1}{4}{N_m}\psi E}}{{{{\left( {\frac{1}{4} - \psi \left( {1 - \rho } \right) E} \right)}^2}}} \\ 
&=\frac{1}{8}\phi \left( - \sqrt {\frac{{{N_m}\psi (1 - \rho )E}}{{\frac{1}{4} - \psi (1 - \rho )E}}} \right)\times \frac{{\sqrt {{N_m}\psi E} }}{{\sqrt {1 - \rho } {{(\frac{1}{4} - \psi (1 - \rho )E)}^{\frac{3}{2}}}}},
\end{split}    
\end{align}
respectively.

We denote $g(\rho ) = f'(\rho )$, then,\\
\begin{align}
\begin{split}
\label{g_rho1}
g(\rho ) &= \frac{1}{4}\phi (\sqrt {\frac{{{N_m}(\frac{1}{2} - \sqrt {\psi \rho E} )}}{{\frac{1}{2} + \sqrt {\psi \rho E} }}} ) \\
& \times \frac{{\sqrt {{N_m}\psi E} }}{{\sqrt {\rho (\frac{1}{4} - \psi \rho E)}  \times (\frac{1}{2} + \sqrt {\psi \rho E} )}}\\
 & - \frac{1}{8}\phi ( - \sqrt {\frac{{{N_m}\psi \rho E}}{{\frac{1}{4} - \psi \rho E}}} ) \times \frac{{\sqrt {{N_m}\psi E} }}{{\sqrt \rho  {{(\frac{1}{4} - \psi \rho E)}^{\frac{3}{2}}}}}\\
 & - \frac{1}{4}\phi (\sqrt {\frac{{{N_m}(\frac{1}{2} - \sqrt {\psi (1 - \rho )E} )}}{{\frac{1}{2} + \sqrt {\psi (1 - \rho )E} }}} ) \\
 &\times \frac{{\sqrt {{N_m}\psi E} }}{{\sqrt {(1 - \rho )} \sqrt {\frac{1}{4} - \psi (1 - \rho )E} (\frac{1}{2} + \sqrt {\psi (1 - \rho )E} )}}\\
 & + \frac{1}{8}\phi ( - \sqrt {\frac{{{N_m}\psi (1 - \rho )E}}{{\frac{1}{4} - \psi (1 - \rho )E}}} ) \times \frac{{\sqrt {{N_m}\psi E} }}{{\sqrt {1 - \rho } {{(\frac{1}{4} - \psi (1 - \rho )E)}^{\frac{3}{2}}}}}.
\end{split}
\end{align}

And, $g(1-\rho)$ can be expressed as

\begin{align}
\begin{split}
\label{g_1rho}
g(1 - \rho ) & = \frac{1}{4}\phi (\sqrt {\frac{{{N_m}(\frac{1}{2} - \sqrt {\psi (1 - \rho )E} )}}{{\frac{1}{2} + \sqrt {\psi (1 - \rho )E} }}} ) \\
& \times \frac{{\sqrt {{N_m}\psi E} }}{{\sqrt {(1 - \rho )(\frac{1}{4} - \psi (1 - \rho )E)}  \times (\frac{1}{2} + \sqrt {\psi (1 - \rho )E} )}}\\
 & - \frac{1}{8}\phi ( - \sqrt {\frac{{{N_m}\psi (1 - \rho )E}}{{\frac{1}{4} - \psi (1 - \rho )E}}} ) \\
 & \times \frac{{\sqrt {{N_m}\psi E} }}{{\sqrt {1 - \rho } {{(\frac{1}{4} - \psi (1 - \rho )E)}^{\frac{3}{2}}}}}\\
 & - \frac{1}{4}\phi (\sqrt {\frac{{{N_m}(\frac{1}{2} - \sqrt {\psi \rho E} )}}{{\frac{1}{2} + \sqrt {\psi \rho E} }}} ) \\
 & \times \frac{{\sqrt {{N_m}\psi E} }}{{\sqrt \rho  \sqrt {\frac{1}{4} - \psi \rho E} (\frac{1}{2} + \sqrt {\psi \rho E} )}}\\
 & + \frac{1}{8}\phi ( - \sqrt {\frac{{{N_m}\psi \rho E}}{{\frac{1}{4} - \psi \rho E}}} ) \times \frac{{\sqrt {{N_m}\psi E} }}{{\sqrt \rho  {{(\frac{1}{4} - \psi \rho E)}^{\frac{3}{2}}}}}.
\end{split}
\end{align}

{\color{black}By utilizing (\ref{g_rho1}) and (\ref{g_1rho})}, we can attain
\begin{align}
\begin{split}
\label{g_rho}
g(\rho ) + g(1 - \rho ) = 0.
\end{split}
\end{align}

It's evident from (\ref{g_rho}) that $f'(0.5) = g(0.5) = 0$, indicating that $\rho = 0.5$ acts as the extremum point of the BER function $f(\rho)$. This implies that the optimal BER performance is attained when $\rho = 0.5$ within the context of the considered cooperative MC. 
\subsection{Optimization scheme for $K>2$}
For $K>2$, it becomes clear that obtaining a closed-form expression for the optimal allocation coefficient $\rho$ is not feasible. As a result, we employ a genetic algorithm to efficiently determine the optimal value of $\rho$. 

The genetic algorithm is a metaheuristic optimization technique inspired by the process of natural selection and genetics. It mimics the process of evolution by iteratively evolving a population of potential solutions to a problem. Through the genetic algorithm, we create an initial population of candidate solutions, each represented as a set of parameters that define the value of $\rho$. These candidate solutions undergo selection, crossover, and mutation operations in each generation. The selection process favors solutions with better fitness, which are more likely to survive and pass their genetic information to the next generation. Crossover and mutation introduce genetic diversity, allowing the population to explore the solution space more effectively. As the genetic algorithm progresses through multiple generations, it converges towards an optimal solution, improving the value of $\rho$ that maximizes the desired objective function, which means that the algorithm achieves better total BER performance. By leveraging the principles of natural selection and genetics, the genetic algorithm provides an efficient and effective way to address complex optimization problems without relying on analytical expressions. By employing the genetic algorithm, we can effectively explore the solution space and find near-optimal energy allocation strategies that minimize the total BER in the multi-user cooperative MC system. The flexibility and adaptability of the genetic algorithm make it a suitable optimization approach for this type of complex and nonlinear problem. The pseudocode can be expressed as

\begin{algorithm}
\SetAlgoLined
\SetKwInOut{Input}{Input}
\SetKwInOut{Output}{Output}
\caption{Genetic Algorithm for Energy Allocation Optimization}

\Input{Energy allocation variables $E_{u_1}, E_{u_2}, ..., E_{u_K}$, the parameters of the transmitter.}
\Output{Optimal energy allocation strategy.}
\BlankLine
\textbf{Step 1: Encoding}\;
Encode the energy allocation variables $\rho$ as genes within a chromosome\;
\BlankLine
\textbf{Step 2: Initialization}\;
Create an initial population of chromosomes randomly\;
\BlankLine
\textbf{Step 3: Fitness Evaluation}\;
Evaluate the fitness of each chromosome based on the objective function $\min \limits_{{E_{u_1}}, {E_{u_2}}, ..., {E_{u_k}}} \sum_{k = 1}^K {P_{e,u_k}}$\;
\BlankLine
\textbf{Step 4: Selection}\;
Select chromosomes for reproduction based on their fitness\;
\BlankLine
\textbf{Step 5: Crossover}\;
Perform crossover between selected parent chromosomes to generate offspring strategies\;
\BlankLine
\textbf{Step 6: Mutation}\;
Introduce random changes into the offspring chromosomes to maintain diversity\;
\BlankLine
\textbf{Step 7: Repeat Steps 3 to 6}\;
Repeat Steps 3 to 6 for a certain number of generations\;
\BlankLine
\textbf{Step 8: Termination and Solution Extraction}\;
Terminate the algorithm based on the stopping criterion: $\sum_{k=1}^K E_{u_k} \leq E_{\text{total}}$; $P_{e,k} \leq \text{BER}_{\text{threshold}}, \forall k = 1, 2, \ldots, K; E_{u_k} \geq 0, \forall k = 1, 2, \ldots, K$.\;
Extract the best chromosome as the optimal energy allocation strategy\;
\BlankLine
\textbf{Step 9: Analysis and Results}\;
Analyze and interpret the obtained results, including the minimum total BER and the optimal energy allocation strategy\;
Perform sensitivity analysis to assess the impact of parameter variations\;
\end{algorithm}

\section{Numerical and simulation results} 
This section explores the BER performance of the cooperative MC system under constrained energy. It investigates the impact of varying numbers of molecules in the reservoirs as well as the number of transmitted molecules. Furthermore, the study delves into the analysis of the optimal energy allocation coefficient $\rho$.
Unless otherwise specified, the simulations were conducted using the following values and parameters: Boltzmann's constant $k = 1.3807 \times 10^{-23}$ J/K,  absolute temperature $T =  298.15$ K, the number of transmitted molecules $N_m = 4\times10^4$, the total energy cost $E = 4\times10^{-16}$ J, the initial number of molecules in the low and high reservoirs $n_L = n_H = 6 \times 10^8$, and the initial concentration of $k_2$ is $c_{k_2}=0.5$.

\begin{figure}[!h]
   \centering
   \includegraphics[width=0.5\textwidth]{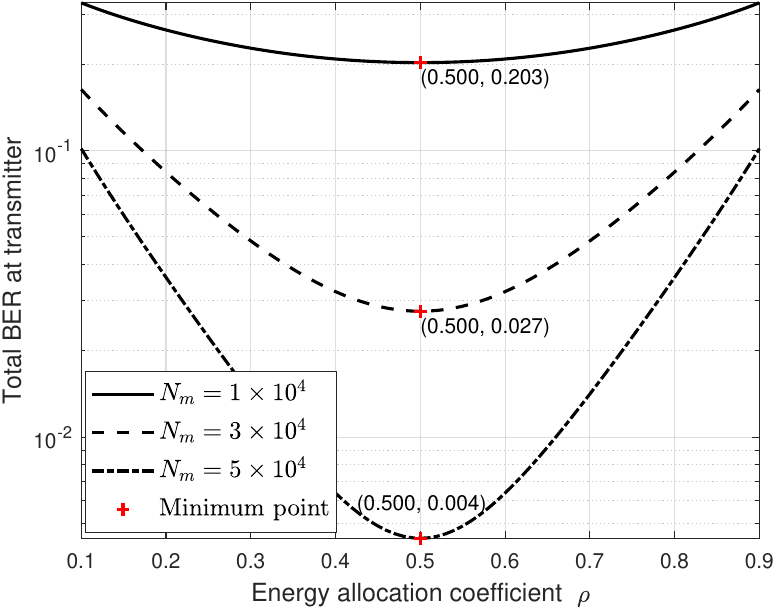}\\
   \caption{\color{black}{The total BER at the transmitter for $u_1$ and $u_2$, where the number of molecules in the reservoirs are both $12\times10^8$.}}\label{ber_rho_n1=n2} 
\end{figure}

\begin{figure}[!h]
   \centering
   \includegraphics[width=0.5\textwidth]{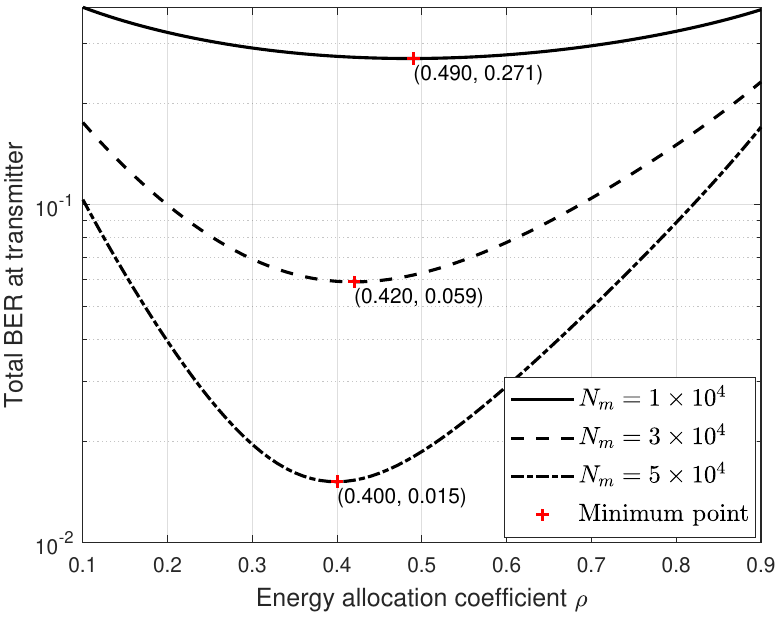}\\
   \caption{\color{black}{The total BER at the transmitter for $u_1$ and $u_2$, where the number of molecules in the reservoirs are $12\times10^8$ and $16\times10^8$, respectively.}}\label{ber_rho_n1_ne_n2} 
\end{figure}

\begin{figure}[!h]
   \centering
   \includegraphics[width=0.5\textwidth]{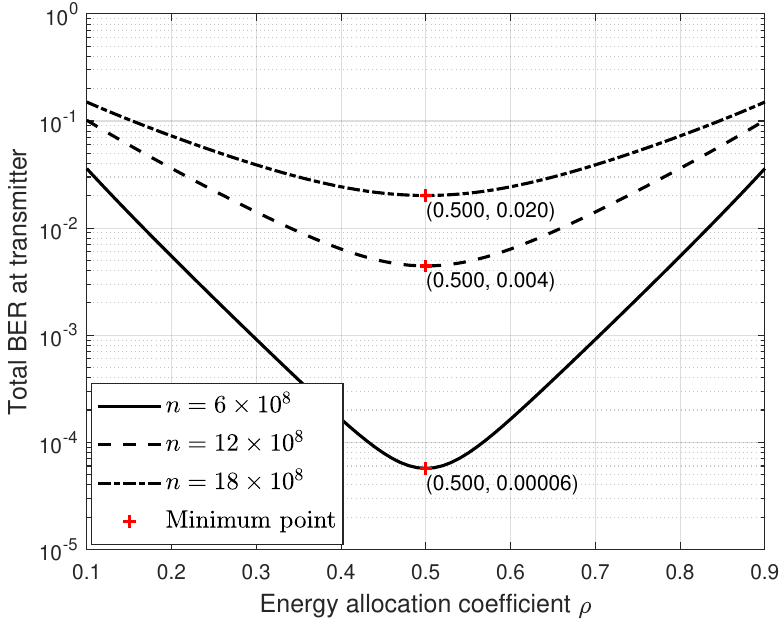}\\
   \caption{The total BER at the transmitter varies with $\rho$ under different numbers of molecules in the reservoirs. The number of transmitted molecules $N_m=5\times10^4$.}\label{ber_rho_n} 
\end{figure}

The study investigates the variation of the total BER with the energy allocation coefficient $\rho$ considering two users across Fig. \ref{ber_rho_n1=n2} to Fig. \ref{ber_rho_n}. As shown in Fig. \ref{ber_rho_n1=n2}, the BER attains its optimal performance at $\rho=0.5$, which aligns with the theoretical analysis in (\ref{g_rho}). This finding indicates that for transmitters of the same size, the best BER performance is achieved when an equal amount of energy is allocated. Additionally, in cases with a larger number of transmitted molecules, there is a higher probability for the transmitted molecules to follow the distribution in the reservoirs, resulting in improved performance. Compared to traditional MoSK, where released molecules are assumed to be pure, and errors arising from the transmitter are not considered, our imperfect transmitter MoSK takes into account the mixture of molecules in the reservoirs, making it impossible to ignore the increase in BER originating from the transmitter.

In Fig. \ref{ber_rho_n1_ne_n2}, we investigate the total BER varies with the energy allocation coefficient $\rho$ under different numbers of transmitted molecules. Unlike Fig. \ref{ber_rho_n1=n2}, the total numbers of molecules for $u_1$ and $u_2$ are different, therefore, the energy allocated to different transmitters is also different, and for a larger number of molecules in the transmitter, it needs to allocate more energy, then, the transmitter can achieve better performance. As shown in the figure, for a larger number of transmitted molecules $N_m$, the BER achieves better performance, meanwhile, the optical allocation coefficient $\rho$ also varies with $N_m$. This is because the distribution of molecules in the reservoirs is different, making the BER performance under different numbers of transmitted molecules different. 

Fig. \ref{ber_rho_n} demonstrates the relationship between the total BER and the energy allocation coefficient $\rho$ for varying numbers of molecules in the reservoirs. As depicted in Fig. \ref{ber_rho_n1=n2}, for the same number of molecules in the users, the optimal BER performance is achieved when $\rho$ is set to 0.5. Additionally, we observe that a smaller number of molecules in the reservoirs results in a better BER performance. This is due to the larger concentration difference between molecules $k_1$ and $k_2$ which enhances the BER performance with the given energy allocation.

\begin{figure}[!h]
   \centering
   \includegraphics[width=0.5\textwidth]{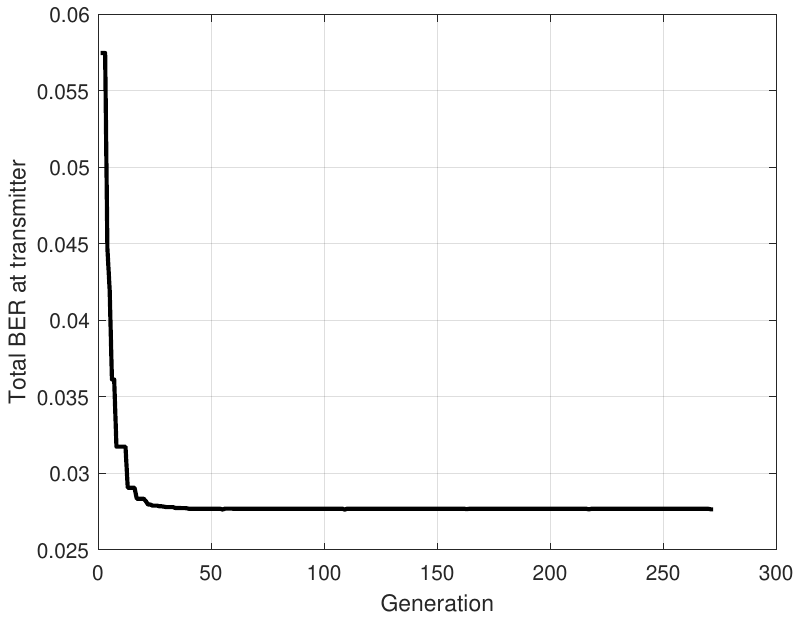}\\
   \caption{The total BER at the transmitter, for $u_1$,$u_2$ and $u_3$, the energy allocation coefficient $\rho$ are 0.20, 0.34, 0.45, respectively. The number of molecules in the reservoirs is $6\times10^8$, $12\times10^8$,$18\times10^8$. the number of transmitted molecules is $5\times10^4$.}\label{ber_rho_gen_ne_n} 
\end{figure}

\begin{figure}[!h]
   \centering
   \includegraphics[width=0.5\textwidth]{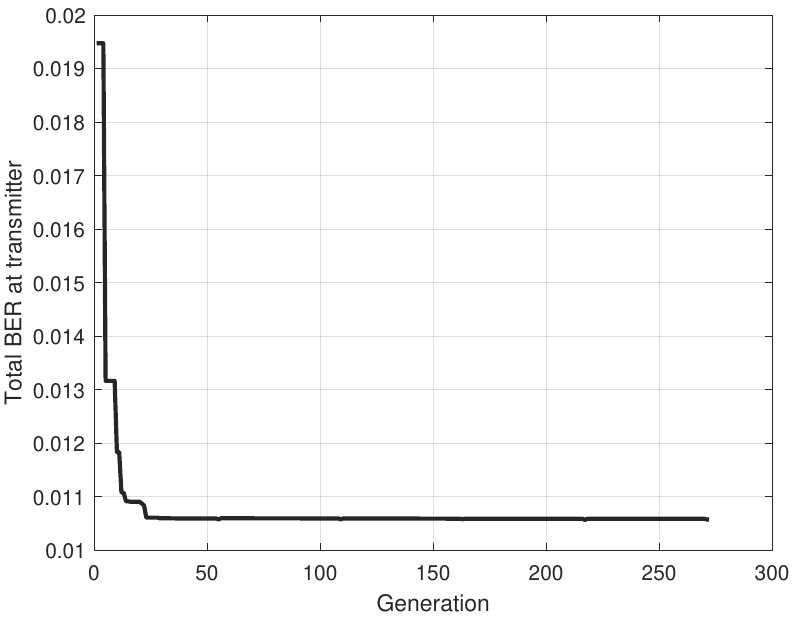}\\
   \caption{The total BER at the transmitter, for $u_1$,$u_2$ and $u_3$, the energy allocation coefficient $\rho$ are 0.49, 0.29, 0.22, respectively. The number of molecules in the reservoirs is $6\times10^8$, the number of transmitted molecules is $2\times10^4$,$4\times10^4$,$6\times10^4$.}\label{ber_rho_gen_ne_Nm} 
\end{figure}

From Fig. \ref{ber_rho_gen_ne_n} to Fig. \ref{ber_rho_gen_ne_Nm}, the BER performance varies with the energy allocation coefficient $\rho$ by employing the genetic algorithm considering three users are investigated. In Fig. \ref{ber_rho_gen_ne_n}, there are different numbers of molecules in the reservoirs for the users are considered, while in Fig. \ref{ber_rho_gen_ne_Nm}, the number of molecules in the reservoirs is fixed, but for a different number of transmitted molecules. In both of these two figures, the BER quickly converges to a stable state. Similar to  Fig. \ref{ber_rho_n1_ne_n2}, as shown in Fig. \ref{ber_rho_gen_ne_n} for different numbers of molecules in the reservoirs, the optical allocation coefficient $\rho$ are also different. However, as shown in Fig. \ref{ber_rho_gen_ne_Nm}, the number of transmitted molecules for different users is different, and the optical allocation coefficient $\rho$ is also different, this is because the number of transmitted also affects the performance of the transmitter, making the $\rho$ varies. Meanwhile, compared with Fig. \ref{ber_rho_gen_ne_n} and Fig. \ref{ber_rho_gen_ne_Nm}, a smaller number of molecules in the reservoirs also make better BER performance. 

\section{Conclusion}
This paper explored a cooperative MC system with imperfect transmitters, featuring low and high reservoirs containing different types of molecules. The system utilized free energy to move molecules from one reservoir to another, creating concentration differences that allowed for information encoding on the molecule types. Due to limited total energy within a given area, the study analyzed the number of molecules moved for a specific energy cost. Furthermore, the paper examined energy allocation strategies among users to achieve improved BER performance. The proposed strategies were rigorously evaluated through theoretical analysis and simulations, demonstrating their effectiveness. The findings revealed a crucial thermodynamic tradeoff between BER and reservoir size, including the number of molecules in the reservoirs and the transmitted molecules. These insights shed light on designing more efficient communication systems. Future research will focus on investigating energy allocation strategies that consider channel parameters, aiming to enhance the overall performance of the cooperative MC system.

\appendix 
\section{}
$P_{e,u_1}$ and $P_{e,u_2}$ defined in (\ref{f_rho}) can be {\color{black}expressed in (\ref{appendix-equation-1}) and (\ref{appendix-equation-2}), respectively, on the next page.}
\begin{figure*}[ht]
    \centering
\begin{align} \label{appendix-equation-1}
&{P_{e,{u_1}}} = \frac{1}{2}\left[ {\left( {1 - {P_{e,{u_1}}}\left( {Y = 0|X = 0} \right)} \right)} \right. + \left. {\left( {1 - {P_{e,{u_1}}}\left( {Y = 1|X = 1} \right)} \right)} \right]\\ \notag
 &= \frac{1}{2}\left[ {1 - \Phi \left( {\frac{{{N_m} - {\mu _0}}}{{{\sigma _0}}}} \right) + \Phi \left( {\frac{{{N_m}(1 - c_k) - {\mu _0}}}{{{\sigma _0}}}} \right) + 1 -  {\Phi \left( {\frac{{{N_m} - {\mu _1}}}{{{\sigma _1}}}} \right) + \Phi \left( {\frac{{{N_m}c_k - {\mu _1}}}{{{\sigma _1}}}} \right)} } \right]\\ \notag
 &= \frac{1}{2}\left[ {1 - \Phi \left( {\frac{{{N_m}{c_{{k_2},L}}}}{{\sqrt {{N_m}{c_{{k_2},L}}\left( {1 - {c_{{k_2},L}}} \right)} }}} \right) + \Phi \left( {\frac{{{N_m}\left( {{c_{{k_2},L}} - c_k} \right)}}{{\sqrt {{N_m}{c_{{k_2},L}}\left( {1 - {c_{{k_2},L}}} \right)} }}} \right)} \right] \\ \notag
 &+ \frac{1}{2}\left[ {1 - \Phi \left( {\frac{{{N_m}\left( {1 - {c_{{k_2},H}}} \right)}}{{\sqrt {{N_m}{c_{{k_2},H}}\left( {1 - {c_{{k_2},H}}} \right)} }}} \right) + \Phi \left( {\frac{{{N_m}\left( {c_k - {c_{{k_2},H}}} \right)}}{{\sqrt {{N_m}{c_{{k_2},H}}\left( {1 - {c_{{k_2},H}}} \right)} }}} \right)} \right]\\ \notag
 &= \frac{1}{2}\left[ {1 - \Phi \left( {\sqrt {\frac{{{N_m}\left( {{c_k} - \sqrt {\frac{{{c_k}}}{{KT{n_{k,L}}}}\rho E} } \right)}}{{1 - \left( {{c_k} - \sqrt {\frac{{{c_k}}}{{KT{n_{k,L}}}}\rho E} } \right)}}} } \right) + \Phi \left( {\frac{{-{N_m}{  \sqrt {\frac{{{c_k}}}{{KT{n_{k,L}}}}\rho E} }}}{{\sqrt {{N_m}\left( {{c_k} - \sqrt {\frac{{{c_k}}}{{KT{n_{k,L}}}}\rho E} } \right)\left( {1 - \left( {{c_k} - \sqrt {\frac{{{c_k}}}{{KT{n_{k,L}}}}\rho E} } \right)} \right)} }}} \right)} \right] \\ \notag
 &+ \frac{1}{2}\left[ {1 - \Phi \left( {\sqrt {\frac{{{N_m}\left( {1 - {{c_k} - \sqrt {\frac{{{c_k}}}{{KT{n_{k,H}}}}\rho E} }} \right)}}{{{c_k} + \sqrt {\frac{{{c_k}}}{{KT{n_{k,H}}}}\rho E} }}} } \right) + \Phi \left( {\frac{{-{N_m} {  \sqrt {\frac{{{c_k}}}{{KT{n_{k,H}}}}\rho E} }}}{{\sqrt {{N_m}\left( {{c_k} + \sqrt {\frac{{{c_k}}}{{KT{n_{k,H}}}}\rho E} } \right)\left( {1 - {{c_k} - \sqrt {\frac{{{c_k}}}{{KT{n_{k,H}}}}\rho E} }} \right)} }}} \right)} \right]. \notag
\end{align}
\end{figure*}
\begin{figure*}[ht]
\centering
\small
\begin{align} \label{appendix-equation-2}
&{P_{e,{u_2}}} = \frac{1}{2}\left[ {\left( {1 - {P_{e,{u_2}}}\left( {Y = 0|X = 0} \right)} \right)} \right. + \left. {\left( {1 - {P_{e,{u_2}}}\left( {Y = 1|X = 1} \right)} \right)} \right] \\ \notag
&=1- \frac{1}{2}\left[ \Phi \left( {\sqrt {\frac{{{N_m}\left( {{c_k} - \sqrt {\frac{{{c_k}}}{{KT{n_{k,L}}}}\left( {1 - \rho } \right)E} } \right)}}{{1 -  {{c_k} + \sqrt {\frac{{{c_k}}}{{KT{n_{k,L}}}}\left( {1 - \rho } \right)E} } }}} } \right) - \Phi \left( {\frac{{-{N_m}\left( { {  \sqrt {\frac{{{c_k}}}{{KT{n_{k,L}}}}\left( {1 - \rho } \right)E} }  } \right)}}{{\sqrt {{N_m}\left( {{c_k} - \sqrt {\frac{{{c_k}}}{{KT{n_{k,L}}}}\left( {1 - \rho } \right)E} } \right)\left( {1 -  {{c_k} + \sqrt {\frac{{{c_k}}}{{KT{n_{k,L}}}}\left( {1 - \rho } \right)E} } } \right)} }}} \right)\right] \\ \notag
&- \frac{1}{2}\left[\Phi \left( {\sqrt {\frac{{{N_m}\left( {1 -  {{c_k} - \sqrt {\frac{{{c_k}{n_k}}}{{2KT{n_{k,H}}}}\left( {1 - \rho } \right)E} }} \right)}}{{{c_k} + \sqrt {\frac{{{c_k}}}{{KT{n_{k,H}}}}\left( {1 - \rho } \right)E} }}} } \right) -\Phi \left( {\frac{{-{N_m}\left( { { \sqrt {\frac{{{c_k}}}{{KT{n_{k,H}}}}\left( {1 - \rho } \right)E} } } \right)}}{{\sqrt {{N_m}\left( {{c_k} + \sqrt {\frac{{{c_k}}}{{KT{n_{k,H}}}}\left( {1 - \rho } \right)E} } \right)\left( {1 -  {{c_k} - \sqrt {\frac{{{c_k}}}{{KT{n_{k,H}}}}\left( {1 - \rho } \right)E} } } \right)} }}} \right)\right]. \notag
\end{align}
\end{figure*}
\bibliographystyle{IEEEtran}
\bibliography{references}

\section{Biography Section}
\begin{IEEEbiography}[{\includegraphics[width=1in,height=1.25in,clip,keepaspectratio]{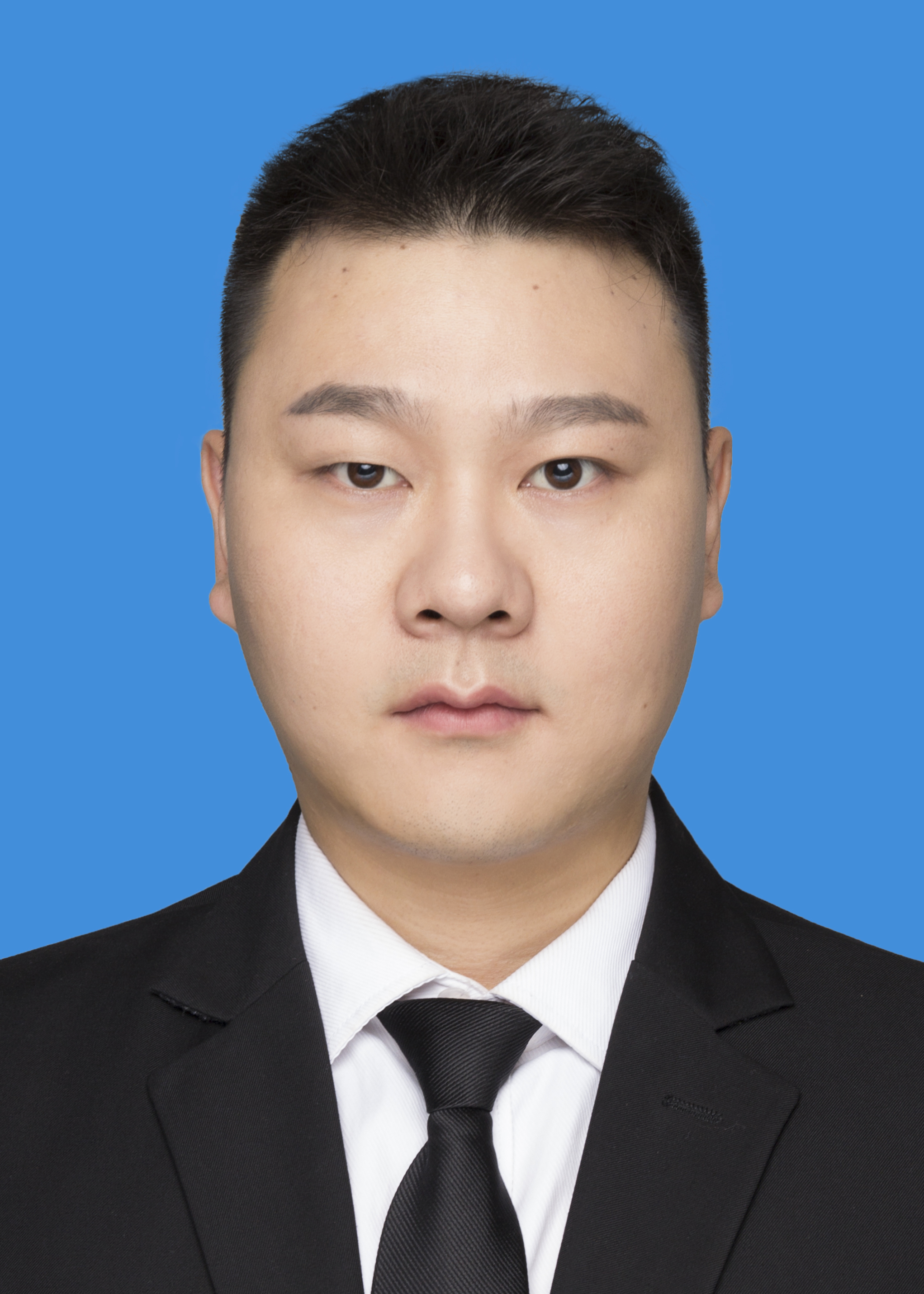}}]{Dongliang Jing}
is a lecturer in the College of Mechanical and Electronic Engineering, Northwest A\&F University, Yangling, China. He received the B.S. degree from Anhui Polytechnic University, Wuhu, China in 2015 and received the Ph.D. degree from Xidian University, Xian, China in 2022. During Nov. 2019 - Nov. 2020, he was a visiting student for molecular communication at York University, Toronto, ON. Canada, under the supervisor of Andrew W. Eckford. His main research interests include wireless and molecular communications.
\end{IEEEbiography}


\begin{IEEEbiography}[{\includegraphics[width=1in,height=1.25in, clip,keepaspectratio]{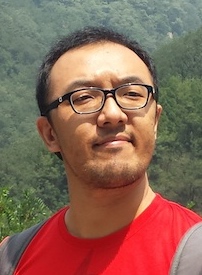}}]{Lin Lin} (Senior Member, IEEE) received B.S. and M.S. degrees in electrical engineering from Tianjin University, China in 2004 and 2007, respectively, and the Ph.D. degree from Nanyang Technological University, Singapore in 2012. He is an Associate Professor with the College of Electronics and Information Engineering, Tongji University, Shanghai, China. He is currently the Chair of IEEE ComSoc Molecular, Biological and Multi-scale Communications Technical Committee, and the Chair of IEEE Nanotechnology Council Nanoscale Communications Technical Committee. He also serves as Associate Editor for IEEE Nanotechnology Magazine and IEEE Access. His research interests include molecular communications, neural communications, internet of nanothings, and body sensor networks.
\end{IEEEbiography}

\begin{IEEEbiography}[{\includegraphics[width=1in,height=1.25in, clip,keepaspectratio]{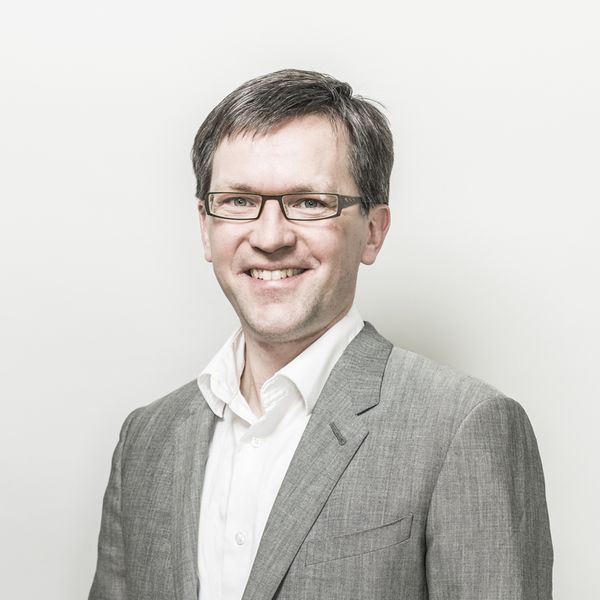}}]{Andrew W. Eckford} (Senior Member, IEEE) is a Professor in the Department of Electrical Engineering and Computer Science at York University, Toronto, Ontario. His research interests include the application of information theory to biology, and the design of communication systems using molecular and biological techniques. His research has been covered in media including The Economist, The Wall Street Journal, and IEEE Spectrum. His research received the 2015 IET Communications Innovation Award, and was a finalist for the 2014 Bell Labs Prize. He is also a co-author of the textbook Molecular Communication, published by Cambridge University Press. Andrew received the B.Eng. degree from the Royal Military College of Canada in 1996, and the M.A.Sc. and Ph.D. degrees from the University of Toronto in 1999 and 2004, respectively, all in Electrical Engineering. Andrew held postdoctoral fellowships at the University of Notre Dame and the University of Toronto, prior to taking up a faculty position at York in 2006. He has held courtesy appointments at the University of Toronto and Case Western Reserve University. In 2018, he was named a Senior Fellow of Massey College, Toronto.
\end{IEEEbiography}

\vfill
\end{document}